# An Intelligent Fire Alert System using Wireless Mobile Communication

Mahdi Nasrullah Al-Ameen
Department of Computer Science and Engineering
University of Texas at Arlington
Arlington, TX, USA
mahdi.al-ameen@mavs.uta.edu

***Abstract*—The system has come to light through the way of inspiration to develop a compact system, based on the fundamental ideas of safety, security and control. Once this system is installed to operation specifying temperature and smoke threshold, in case of any emergency situation due to increasing temperature and/or smoke at place surpassing the threshold, the system immediately sends automatic alert-notifications to the users, concerned with the situations. The user gets total control over the system through mobile SMS, even from the distant location, that to change the threshold, turn on/off the feature of sending 'alert notification' and also to reset the system after the emergency situation is overcome. Before executing any command (through SMS) from the user, the system asks for the preset password to verify an authorized user. The security issues have been considered with utter attention in this system to ensure its applicability in industries and business organizations, where security is an important concern. Hence, the fundamental ideas of safety, security and control have been entirely ensured through the system, which have definitely worked as the gear moving factor to look for a new dimension of an 'Intelligent Fire Alert System'.**

***Keywords-Temperature sensor; smoke sensor; microcontroller; mobile-communication; AT command.***

## I. INTRODUCTION

Temperature and smoke sensing alert system is motivated to sense the temperature and smoke and send the alert in an intelligent fashion in case of emergency situation due to fire blow. In every country in the world the fire alarming system is considered to be essential for lots of physical structures including industries, shopping malls and private houses. To note, every year in the United States, over 400,000 residential fires result in 4,000 fatalities and 20,000 injuries. Over 50% of those fatalities occur in homes without temperature and smoke detectors.

To sense the fire-blow with greater sensitivity, the incorporation of the temperature and smoke sensors in a single module is truly essential. Considering the fact my system is a unique one that has not only incorporated multiple number of both type of sensors but also facilitated the service to send SMS and make Call to a number of remote users in case of fire blow. Security is provided through the password and the authorized users can control the system from the distant location by using mobile communication network. And that has made my system eligible for the modern world, where the people can be anywhere at anytime but like to have the control in hand.

Thus the system can play an important role in saving human-lives, ensuring the growth of the economy by saving lots of industries from devastating fire-blow and relieves the people that, in case of any emergency at their home/industries due to fire blow they will be instantly informed, wherever they are.

## II. FEATURES OF THE SYSTEM WITH TECHNICAL OVERVIEW

- Multiple temperature and smoke sensors are incorporated in the system to cover a wide range of area, which are connected to a center microcontroller.

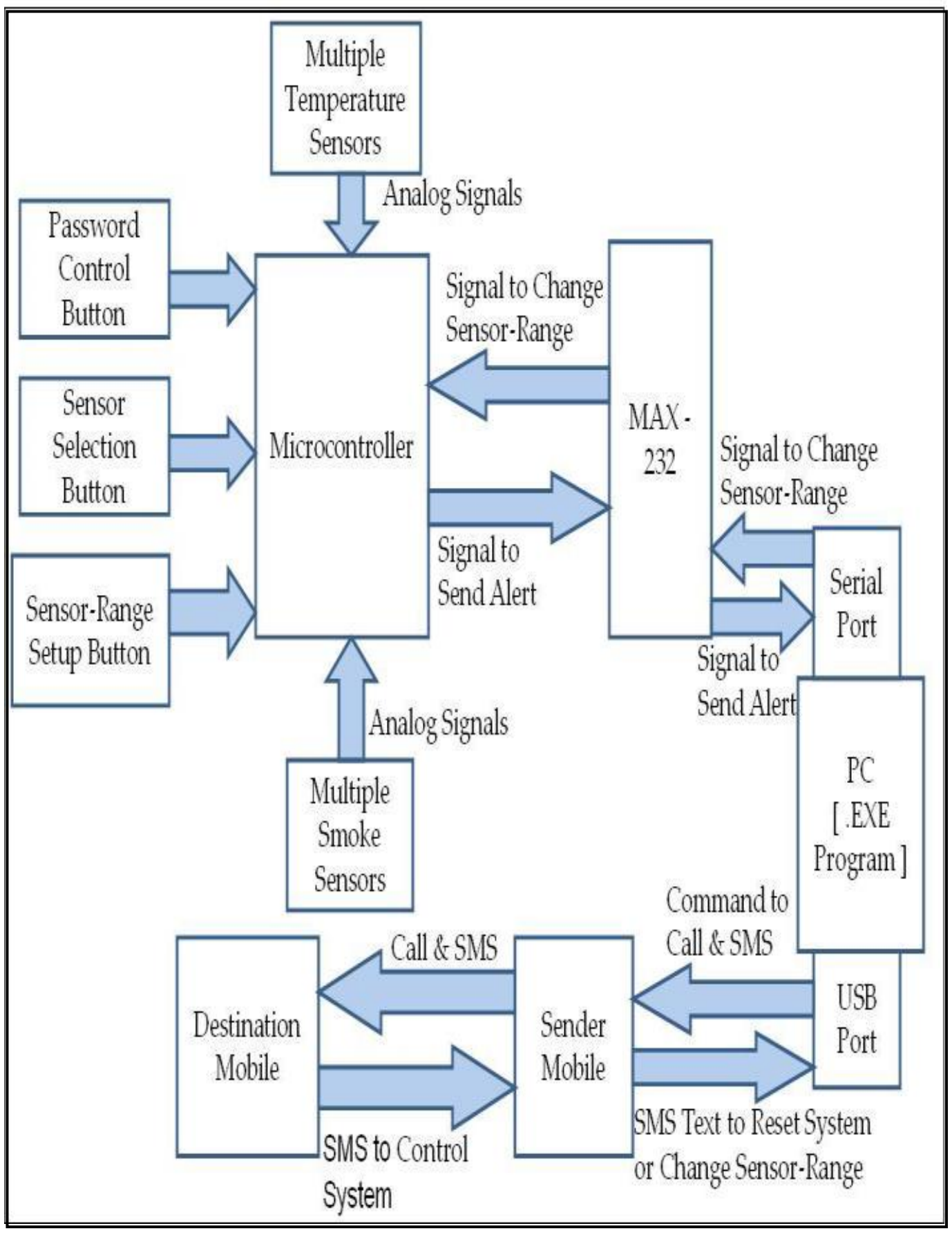

Figure 1. Block diagram of the system

- The user has provision to set distinct threshold for each of the temperature and smoke sensors by pressing corresponding buttons after providing the correct password through password buttons. Threshold of a sensor can be changed from a remote place by using SMS but the correct password must be provided with the new threshold value to have the system execute this instruction.
- The user can change the password by pressing corresponding button, incorporated in the system and the password change mode will hold for sufficient period of time, within which if the password is not set, the system will restart working on the previous password and coming out of the password change mode, will continue normal operation. So the system is intelligent enough to come out of the password change mode and will never stuck to this mode even if the user forgets to come out of the mode by pressing corresponding button. Correct password must be provided by the user to enter password change mode.
- Analog signals from sensors are converted to digital value by A/D converter inside the microcontroller and compared with the preset threshold. When one of the sensors senses temperature/smoke above the threshold, the microcontroller sends a signal to the serial port of a server computer. A program is running on the computer which continuously checks the serial port and on getting the signal from microcontroller it sends appropriate AT command through USB port to the server mobile for broadcasting alert notification through CALL and SMS. The SMS text carries sensor id that has sensed the temperature/smoke above the threshold. So from any remote distance the user can easily identify the specific sensor and the corresponding location where the threshold is surpassed.
- The alert notification can be sent to multiple users and the corresponding mobile numbers can be set/reset by the user in the server mobile through the user interface of the program, running on the computer.
- After the emergency situation is handled; the user can reset the system from remote distance by sending SMS.

To facilitate the mentioned features required hardware system and software programs have been developed and the details are described in next sections.

## III. OVERVIEW OF HARDWARE COMPONENTS WITH CORRESPONDING FEATURES

### *A. Temperature Sensor*

LM 35 is used as temperature sensor. It has an output voltage that is proportional to Celsius temperature. Its scale factor is .01V/°C and it does not require any external calibration or trimming. It maintains an accuracy of +/-0.4 °C at room temperature and +/- 0.8 °C over a range of 0 °C to +100 °C. Another important characteristic of the LM35 is that it draws only 50 micro amps from its supply and possesses a low self-heating capability. Sensor self-heating causes less than 0.1 °C temperature rise in still air [4].

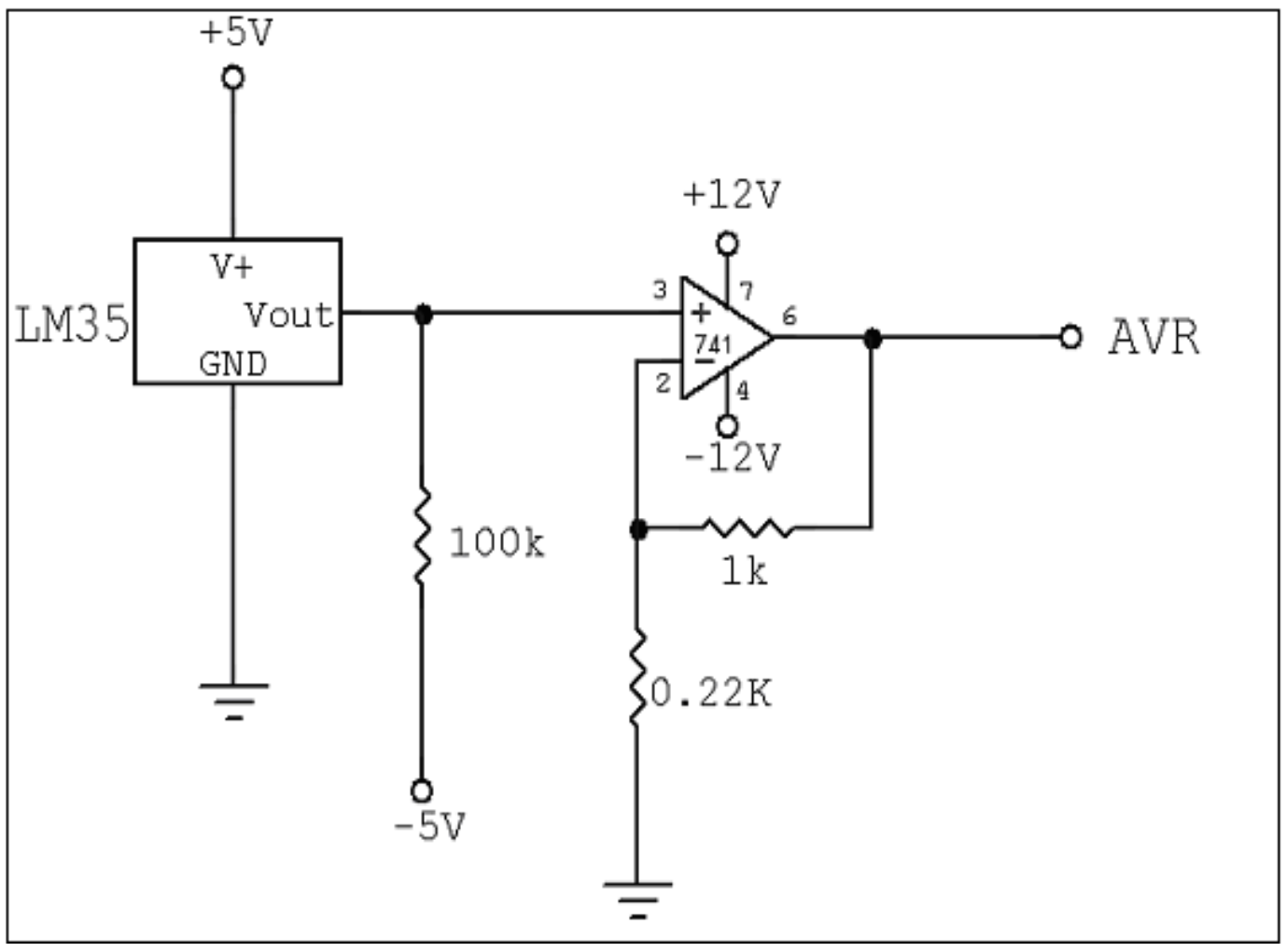

Figure 2. Circuit Diagram of temperature sensor

In my system, the temperature sensor sends an analog signal to the microcontroller which is proportional to the room temperature. To get 50 micro amperes current at the $V_{out}$ pin I have calculated R1 with a negative voltage source.

$$R_1 = \frac{-V_s}{50\mu A}$$

$$= \frac{-(-5V)}{50\mu A} = 100K$$

The output voltage of LM35 is 10mV/°C. Hence I amplify the output voltage of LM35 to get enough voltage in AVR microcontroller's ADC input port. Here the amplifying factor of the non-inverting amplifier is:

$$AF = 1 + \frac{983\Omega}{220\Omega} = 5.47$$

To note, at 30°C temperature the output of LM35 is found 0.30V. After amplifying the output I get $V = 0.3 * 5.47 = 1.641 Volt$ at the microcontroller's input port (ADC0-pin 40 for temperature sensor: 1, ADC1-pin 39 for temperature sensor: 2).I have set 55°C as the default threshold temperature (user can configure the threshold) and at this temperature microcontroller gets:

$$Vtemp_threshold = 55°C * 10mV * 5.47$$

$$Vtemp_threshold = 3.0085V$$

So, at the default threshold temperature, when the microcontroller receives more than 3V at pin: 40 or pin: 39, it sends alert to the server computer's Serial Port to send SMS and make call.

### B. *Smoke Sensor*

The premise of the smoke sensor is a T-shaped chamber with an infrared LED that emits a beam of light across the horizontal portion of the chamber. The left opening of the chamber is for the LED; the right opening, for smoke particles to enter the chamber; and the base opening, for the LDR (Light Dependent Resistor). The chamber itself is made of PVC plastic. The infrared LED is positioned at the left of the chamber; the LDR at the vertical base of the T. The resistance of the LDR is about 430MΩ in dark and is about 11KΩ in bright light.

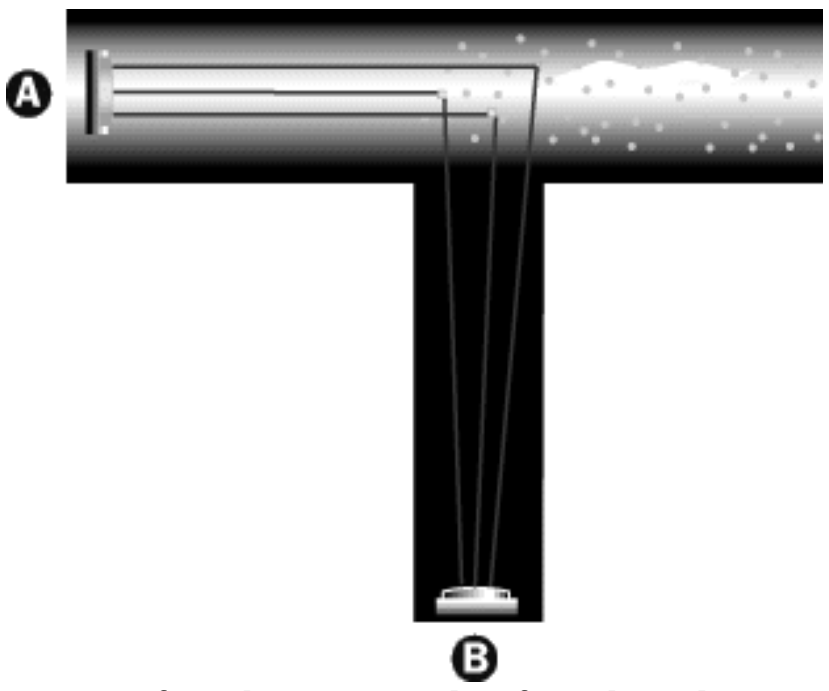

Figure 3. Illustration of working principle of T-shaped pipe, LDR (B) and IR LED (A)

If there is no smoke present inside the chamber, the beam of light is emitted across the top of the T-shaped chamber. Thus, the resistance of the LDR becomes very high as no light has hit it. For this, in the voltage divider circuit maximum voltage drop occurs across the LDR. When smoke particles enter the T-shaped chamber (figure 3), however, the beam of light is scattered by smoke particles. Subsequently, some of the light is directed down into the vertical portion of the T-shaped chamber and strikes the LDR. Thus the resistance of the LDR decreases. So, in the voltage divider circuit the maximum voltage drop occurs across the 9MΩ.

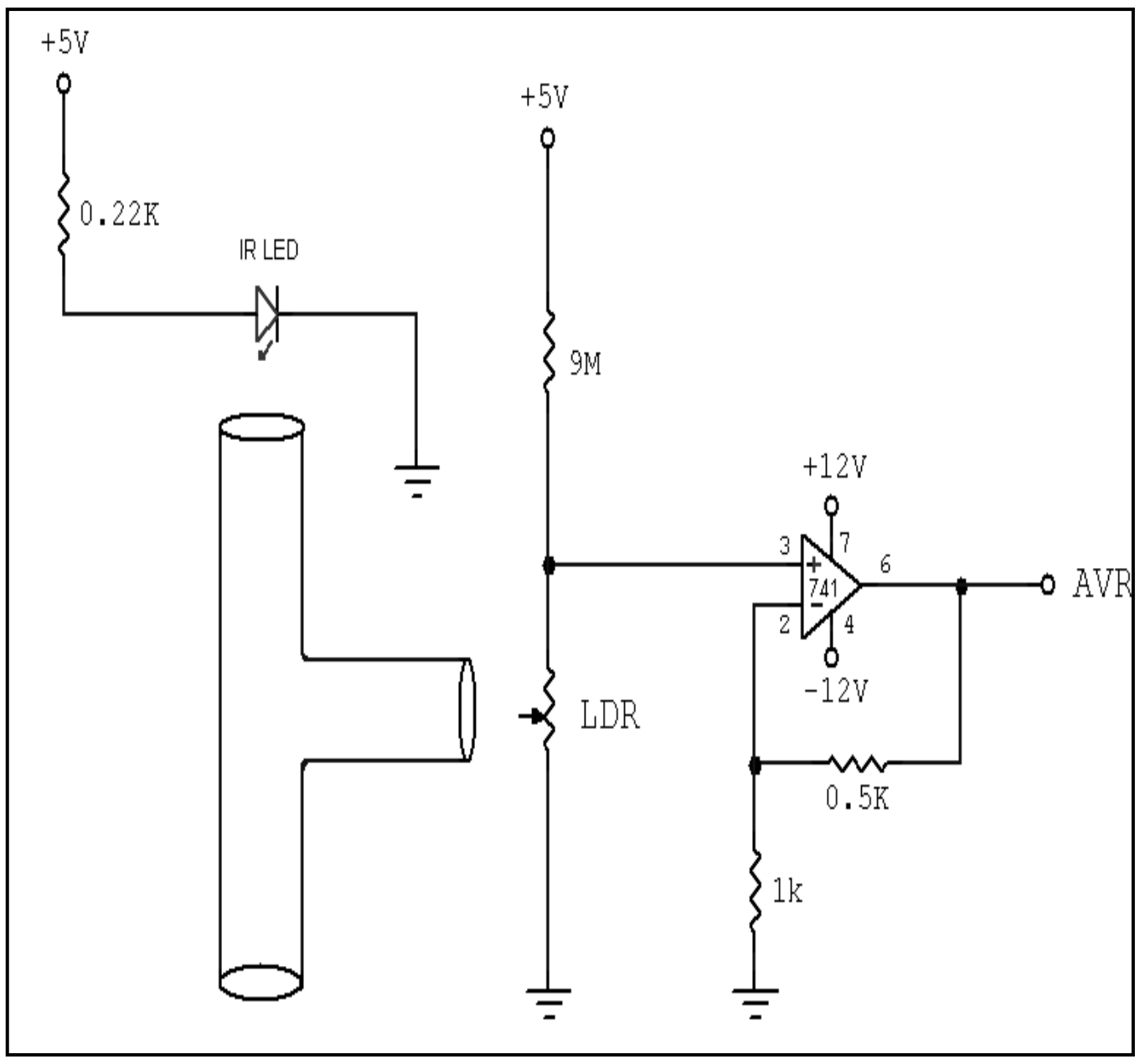

Figure 4. Circuit diagram of smoke sensor

To make the sensor more sensitive I amplify the voltage drop across LDR. The amplified output voltage of smoke sensor: 1 goes as input to Pin 38(ADC2) of the AVR microcontroller and the amplified output voltage of smoke sensor: 2 goes to Pin 37(ADC3). Here the amplifying factor of the non-inverting amplifier is –

$$AF = 1 + \frac{0.5K\Omega}{1K\Omega} = 1.5$$

Through the experiment it has been found that when there is no smoke then the output voltage is about 5.5V. As the intensity of the smoke increases the output voltage gradually decreases and becomes 3.0V-3.5V at a certain level. Thus, for this smoke sensor, I have set 3.0V as the default threshold voltage. So at the default threshold voltage for smoke, when the AVR microcontroller gets less than 3.0V at pin: 38 or pin: 37, it sends alert to server computer’s Serial Port to send SMS and make Call.

### C. *Microcontroller*

In my system Atmega32 (A microcontroller of AVR Large-Series of ATMEL) has been used. It is a high-performance, low power 8bit microcontroller, which has 32K Bytes of in-system self programmable flash memory and 32x8 general purpose working registers with 2K bytes internal SRAM. It has Programmable Serial USART (Universal Synchronous Asynchronous Receiver Transmitter) with 32 programmable I/O lines. Its operating voltage is 4.5 - 5.5V and speed grades are 16 MHz [3].

I have used crystal oscillator of 11.0592 MHz as external clock source (implied to XTAL1 and XTAL2 pin) with 18 pf capacitor for finer clock input to ATMEGA32. For ADC conversion the value of reference voltage I have used is 5V, implied to the AVCC pin and pin no. 31 is needed to be grounded for finer ADC conversion. Pin. 14 (RXD) and pin. 15 (TXD) are connected to Max-232 for serial communication.

### D. *Rs-232 and Max-232*

For serial communication with the server computer’s Serial Port, Rs-232 is needed for interfacing the microcontroller to the server computer. In this event, Rs-232 communicates on EIA-232D standard where +5V to +15V is considered as negative logic voltage and -5V to -15V is considered as positive logic voltage at the transmitting end and at the receiving end +3V to +15V is considered as negative logic voltage and -3V to -15V is considered as positive logic voltage. Due to greater noise margin than TTL logic, reliability in the event of communication through Rs-232 is increased [13].

In my system, pin.7 (RTS) of Rs-232 is directly connected to pin.8 (CTS) and pin.6 (DSR) is directly connected to pin.4 (DTR) where as pin.2 (Rx) and pin.3 (Tx) are connected to the respective pin of Max-232 (described later). A common ground is used for microcontroller, Rs-232 (pin.5 is connected to the ground) and Max-232 to ensure reliable communication.

While data is being transmitted between microcontroller and the server computer’s Serial Port through Rs-232, Max-

232 performs the operation of voltage shifting between TTL logic(used by microcontroller) and EIA-232D standard (used by Rs-232). In fact Max-232 is a dual driver/receiver that includes a capacitive voltage generator to supply TIA/EIA-232-F voltage levels from a single 5-Vsupply. Each receiver converts TIA/EIA-232-F inputs to 5-V TTL/CMOS levels [5].

To note, Max-232 has internal charge pumper that provides output at the range from 7V to 10V at pin 2 (experimentally achieved data), and -7V to -10V at pin 6; depending on the source voltage at pin.16 and the value of the capacitors, used.

### *E. Server computer's serial port and USB port*

The term "serial port" usually identifies hardware more or less compliant to the RS-232 standard. The Microsoft MS-DOS and Windows environments refer to serial ports as COM ports: COM1, COM2, etc. The port speed and device speed must match, though some devices may automatically detect the speed of the serial port [13]. In my system, serial port of computer operates with speed: 115200bps, data bits: 8, stop bit: 1, parity bit: none, flow control: none.

In my system when data cable of the server mobile is connected to the USB port an usb to serial driver relavent with the mobile has to be installed and then the server computer detects the USB port as a Virtual COM port. Identification for the virtual COM port is shown by default, i.e. COM3/COM4 etc. But through the experiment I have come to the point that to ensure the connectivity of the mobile with the USB (Virtual COM Port) along with the reliable serial communication, the identification number of the Virtual COM Port has to be set to COM15 from the computer management wizard of the server computer. I have selected the following configuration for communication with the server through the USB Port (Virtual Com Port): Speed: 9600bps, Data Bits: 8, Stop Bit: 1, Parity: None, Flow Control: None.

### *F. Server mobile*

In my system Nokia 3220 is used as the server mobile, which is nokia series-40 model with internal GPRS modem supporting AT command and conveniently works in text mode instead of PDU mode to send or receive SMS through AT command [11].

### *G. Password mode button*

This button is used to enter the password change mode and the password has to be changed within a specific period of time so that the changed password takes affect on the subsequent operations; othewise the previous password will be in affect. It is connected to PC7 (pin no.29) of Atmega32.

### *H. Password button*

7-buttons are used in my system to set the password. If a button is pressed it stands for 1; otherwise it's 0. So, 128 different combinations of passwords are possible in the system. To set a new password or to change the threshold of a sensor, current combination of the password must be provided by pressing the relevant password buttons. A default password (0x3F) is given in my system which can be changed by the user. Password buttons are connected to PC0( Pin no.28) to PC6 (Pin no.22) pins of Atmega32.

### *I. Sensor selection button*

To change the threshold for a sensor, a specific sensor is selected by the sensor select button and thus distinct range of threshold can be set for each of the sensors. 3-buttons are reserved for sensor selection and thus maximum 8 number of sensors can be incoporated in the system. In the system I have developed 4 sensors have been used and thus 1 sensor selection button has been kept reserved. Sensor selection button 1, 2 and 3 are connected respectively to pin 16, 17 and 18 of Atmega32.

| Button:2 | Button:1 | Selected Sensor |
|---|---|---|
| Not Pressed | Not Pressed | Temperature Sensor:1 |
| Not Pressed | Pressed | Temperature Sensor:2 |
| Pressed | Not Pressed | Smoke Sensor:1 |
| Pressed | Pressed | Smoke Sensor:2 |

TABLE I. SENSOR SELECTION BUTTON FOR TEMPERATURE SENSOR

### *J. Temperature and smoke sensors' threshold-range selection button*

The user can change the default threshold for each temperature and smoke sensor by the corresponding threshold-range select button, which are connected to the port: B of the microcontroller.

| Button Id | Connected Pin No. of Atmega32 | Selected Density-Range For Smoke Sensor |
|---|---|---|
| 1 | PB5 (pin.6) | High |
| 2 | PB6 (pin.7) | Medium |
| 3 | PB7 (pin.8) | Low |

TABLE II. THRESHOLD-RANGE SELECTION BUTTON FOR TEMPERATURE SENSOR

| Button Id | Connected Pin No. of Atmega32 | Selected Range For Temperature Sensor |
|---|---|---|
| 1 | PB0 (pin.1) | >=35 °C |
| 2 | PB1 (pin.2) | >=45 °C |
| 3 | PB2 (pin.3) | >=55 °C |
| 4 | PB3 (pin.4) | >=65 °C |
| 5 | PB4 (pin.5) | >=75 °C |

TABLE III. THRESHOLD-RANGE SELECTION BUTTON FOR SMOKE SENSOR

### *K. LED display*

Led display is used to display the current state of the system (regarding password change) to the user. In this case, when the LED (connected to pin no. 21 of atmega32) glows it indicates the system is currently in password change mode.

The glowing of LED (connected to pin no. 19 of atmega32) indicates that the password is not changed and when the LED (connected to pin no. 20 of atmega32) glows it indicates that the password is successfully changed.

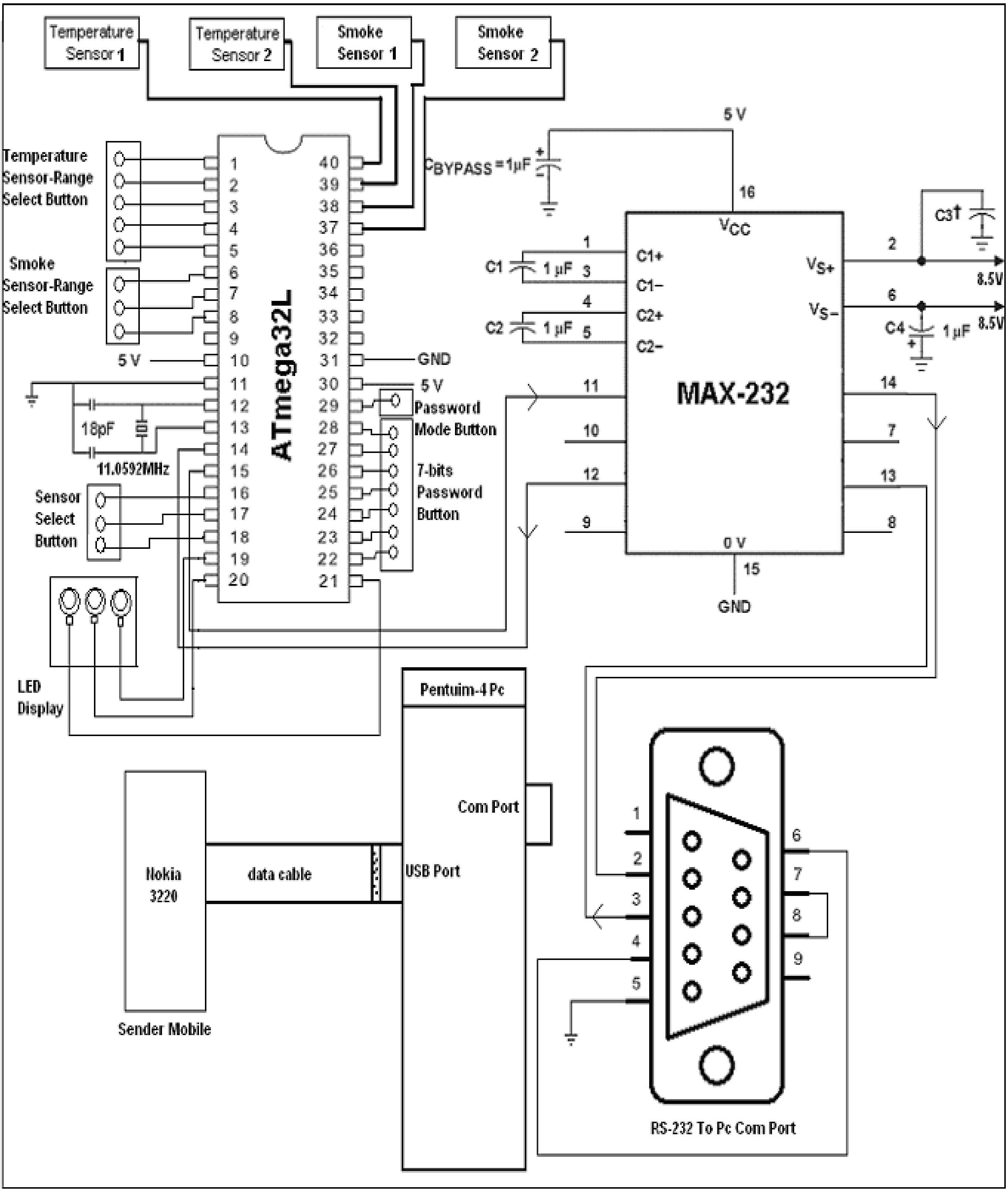

Figure 5. Detailed pin-level diagram of the system

## IV. DESCRIPTION OF THE SOFTWARE PROGRAM

The software program for the system is discussed in two sections.

- First section considers the program written in C# language on .Net environment for the server computer.
- Second section considers the program written in C language for the microcontroller.

### A. *Program for the server computer*

In the software program for the server computer, two threads work in parallel which are carefully maintained to ensure continuous checking of the serial port and successful two-way communications through the USB port. In this case the initialization phase of the program performs the following operations.

#### 1) *Initialization phase*

- Creates an object of type *SerialPort* to communicate with microcontroller through COM1 port at 115200bps with 1 stop bit and no parity bit.
- Creates another object of type *SerialPort* to communicate with server mobile through the Virtual COM Port*: COM15* at 9600bps with 1 stop bit and no parity bit.
- Initialize the server mobile to work on Text Mode while sending and receiving SMS by using AT Command.
- Initialize the server mobile so that its SIM beomes its default memory, where received SMS will be stored by using AT Command.

#### 2) *Detail discussion on the operations performed by the threads*

Two threads are created; one works through COM1 port and another works through COM15 port. At regular interval these threads wake-up automatically and checks if there's any received data in the buffer of the respective COM port or if there is any received data that is read and comes to the program environment; if there is no such data the threads go to sleep again.

Using the parallelism of the thread, execution-program has been developed for each thread. In this case, thread for COM1 checks at regular interval if there's any signal received from the microcontroller. When it is received, it checks the signal (character) and determines for which sensor-output the signal appears (due to the surpassing of the threshold of that sensor). Consequently the program selects a specific memory location of the server mobile by using AT Command to send the SMS text, stored at that location which carries the id and location information of that specific sensor. So, it is prominent that, distinct SMS text is stored initially in different memory location of the server mobile for each sensor.

When a sensor gives the output to the microcontroller after its threshold is surpassed, microcontroller continuosly sends signal to the server computer because it is needed to ensure that, to each of the destination mobiles at least one SMS will be sent for a particular emergency situation.

The second SMS and CALL is sent by the thread for COM15. This thread sends a command to receive SMS from the server mobile at regular interval. User can change the threshold of the sensor at any time and *reset* the system after an emergency situation through this SMS. When a SMS is received the thread checks the command in the SMS and the relevant signal is sent to the microcontroller to change the threshold of a particular sensor. After an emergency situation, the thread can reset the system if the corresponding command is received from the user through SMS.

To note, in my system the microcontroller keeps on working even after an emergency situation occurs to sense the surpassing of threshold of any other sensor. So the program in the server computer is needed to be reset after an emergency situation to send SMS and make Call in case of any other emergency situation; and provision is provided for the user to do it by sending SMS from any remote distance.

In the program, count variable has been used to introduce delay during program execution which is important because of the requirement of having a definite time-gap between the executions of two AT Commands. The delay, maintained in my system is calculated from experimental data.

#### 3) *AT commands used in the software program*

To read a received SMS, following AT commands are used.

- AT+CMGF=1 <Enter>: selecting text mode [12].
- AT+CPMS="SM" <Enter>: selecting the memory of mobile's sim card to save the received SMS [12].
- AT+CMGR=id <Enter>: reading the received SMS. In this command, 'id'= location-id of the inbox, where the received SMS is saved [12].

To send a SMS, required AT commands are as follows:

- AT+CMGF=1 <Enter>: selecting text mode [12].
- AT+CMSS=n,**"** xxxxxxxxx**"** <Enter>: sending SMS. In this command, 'n'=location id of the phone-memory. Usually it denotes the memory location of the outbox. But it may be the memory location of saved items, too. In fact it varies for mobiles of different models and manufacturers. In the mobile used in my system, it denotes the memory location of the outbox. The value of n may be 1, 2, 3…etc. Here, xxxxxxxxx=destination mobile number [12].

To delete a SMS the following AT command is used.

AT+CMGD= id<Enter>: here, 'id'= location-id of the memory from where the SMS is to be deleted [12].

The AT command, used to make a CALL is as follows.

ATD xxxxxxxxx; <Enter>: In this command, xxxxxxxxx= destination mobile number [12].

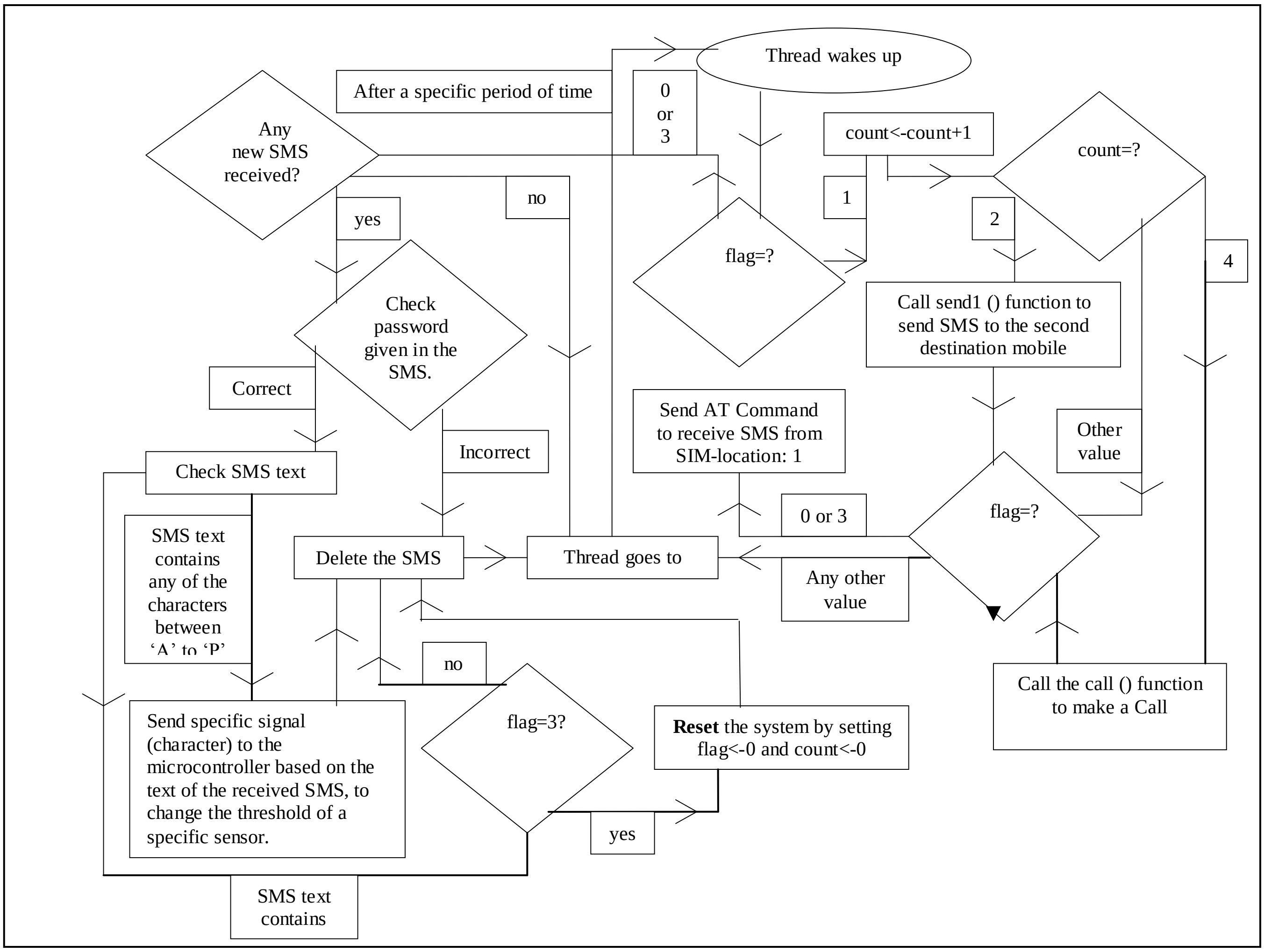

Figure 6. Flow chart for the operation of the thread working through COM15

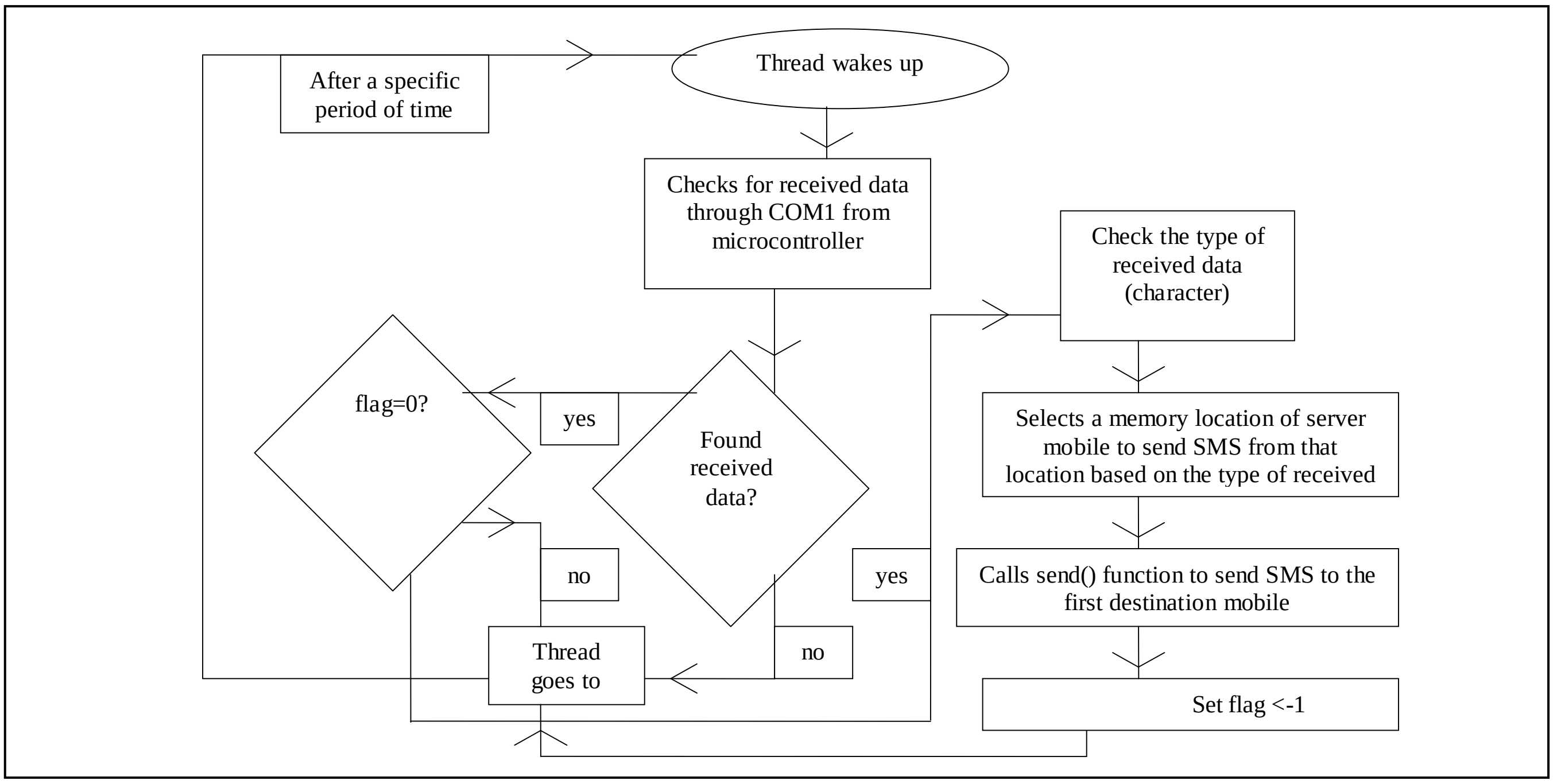

Figure 7. Flow chart for the operation of the thread working through COM1

### B. *Software program for the microcontroller*

The program for the microcontroller is written in a way so that the microcontroller sends distinct alert signal (character) to the server computer for each sensor when the corresponding threshold of that sensor is surpassed. After sending an alert signal the microcontroller continues its regular operation so that it can send alert signal again if the threshold of any other sensor is surpassed.

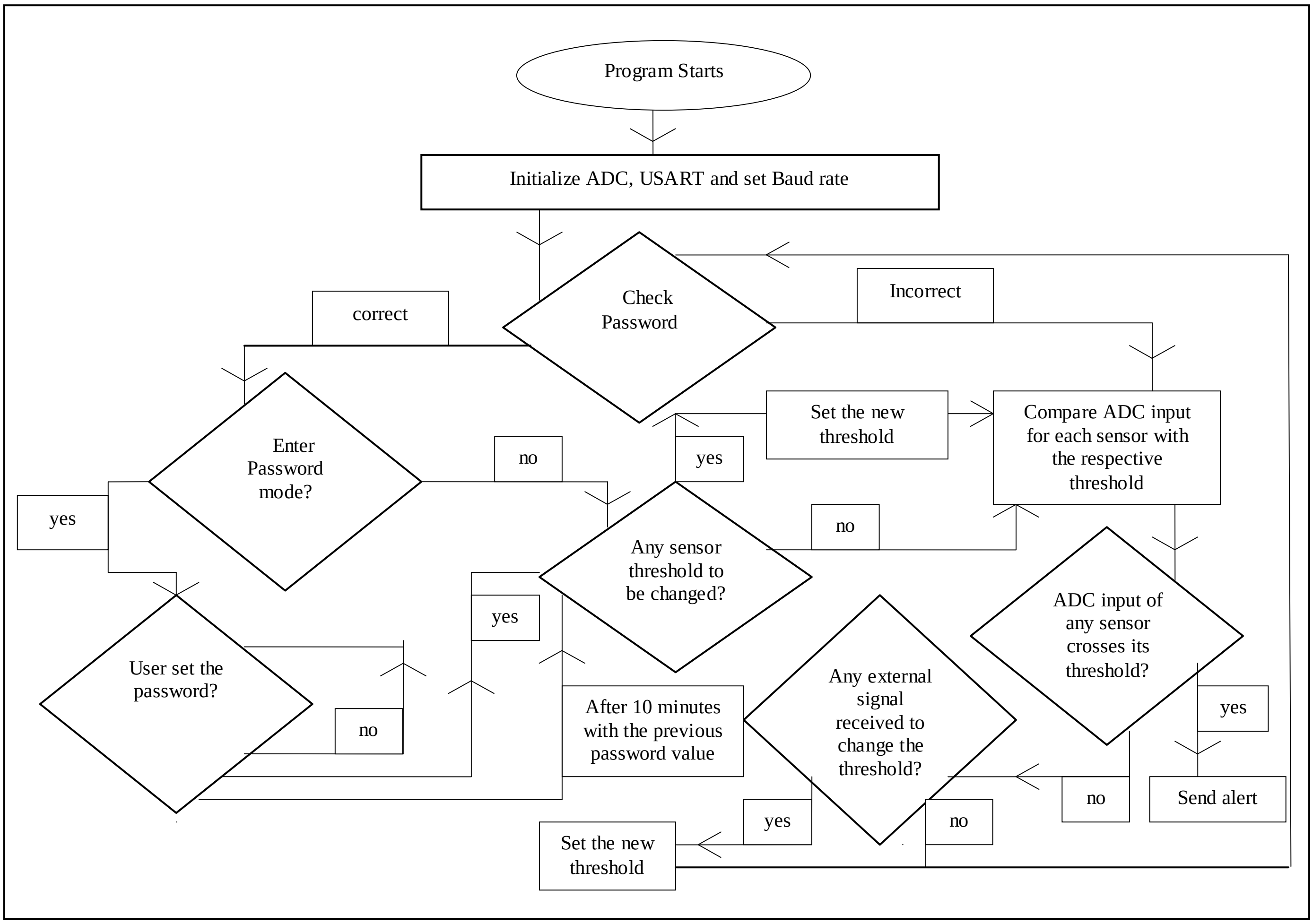

Figure 8. Flow chart for the operation of the microcontroller

## V. DISCUSSION ON SETUP AND MAINTENANCE OF THE SYSTEM

### A. *Setup of the system*

Tasks that are needed to perform for the initial setup of the system are as follows.

- Multiple sensors are incorporated in my system that should be to be set-up at different locations within a physical architecture. The microcontroller and the server mobile are incorporated in a distinct/central module that has to be placed in a safe region, so that in case of any emergency the central module can send alert to the user.
- Physical connections between the sensors and the central microcontroller have been provided in my system and the user has to be careful that the connections are not adversely affected by any means.
- A server computer is needed to be integrated to my system. The user has to ensure that the program (provided in portable CD with my system) for the computer has been properly installed and the connection between the server computer and microcontroller is properly made through the serial port of the computer. Through the USB Port, the computer has to be connected with the server mobile in my system.
- After ensuring the connection between the server mobile and the USB Port of the computer the user has to set the virtual com port identification to COM15 from the computer management wizard of the computer.

- Power connections to the sensors, server computer and the central microcontroller have to be ensured.
- Server mobile has to be charged for minimum 12 hours before starting the sytem for the first time.

After the successful initial set-up, the system is ready to fuction just after few steps:

- Power has to be supplied to the sensors, central microcontroller and the server computer.
- The installed program in the computer has to be started.
- The destination mobile numbers have to be provided to the program interface to send SMS and make Call.
- A password has to be provided to the program interface to control the system from remote distance through the SMS in a secured way.
- The default password has to be provided to the microcontroller by pressing respective buttons. Default password is 1111111. So all the password buttons have to be pressed initially because unpressed buttons stand for 0 and pressed buttons stand for 1. Then the user can give his own password and set the the threshold for each of the sensors. Details on password and threshold setting have been already discussed. Passwords provided to the program in the computer and the microcontroller need not be the same. These two passwords are handled in distinct ways.

Each time the microcontroller is started by supplying power the above steps have to be followed to properly initiate the sytem.

### *B. Maintenance of the system*

Tasks that are needed to perform regularly for the proper maintenance of the system are as follows.

- All the physical connections between different parts of the system should be regularly checked.
- The server mobile has to be regularly charged.
- The user has to be careful that the virus does not adversly affect the installed program in the server computer.
- Every day, for a period of time the system should be relieved from its regular operation to cool down in order to ensure effective performance for a long period of time.

## VI. DISCUSSION ON CONTROL OVER THE SYSTEM

### *A. Setting new threshold and password*

To set a new threshold for a sensor, the user must provide the correct password anvd then select a sensor by pressing respecive buttons to set a new threshold for that sensor. The default threshold for temperature sensor is '55 degree celcius' and 'high density' for the smoke sensor.

To set a new password the user must provide the current password correctly and then press the password mode button to enter the password change mode. Password mode holds for ten minutes. Within this period the new password has to be set. Otherwise the previous password will remain effective.

### *B. Controlling the system from remote distance*

To control the system from remote distance by sending SMS, the format of the SMS text is: xxxxxxxxxx C

Here, xxxxxxxxxx: Maximum ten character long password.

C: A command character to control the system.

In this case, the provided password in the SMS text must match the password, provided through the user interace in the server computer. Othewise the command character will not be effective to control the system.

| **Command Character** | **Command To Execute** |
|---|---|
| A | Set the threshold of the temperature sensor:1 to 35°C |
| B | Set the threshold of the temperature sensor:1 to 45°C |
| C | Set the threshold of the temperature sensor:1 to 55°C |
| D | Set the threshold of the temperature sensor:1 to 65°C |
| E | Set the threshold of the temperature sensor:1 to 75°C |
| F | Set the threshold of the temperature sensor:2 to 35°C |
| G | Set the threshold of the temperature sensor:2 to 45°C |
| H | Set the threshold of the temperature sensor:2 to 55°C |
| I | Set the threshold of the temperature sensor:2 to 65°C |
| J | Set the threshold of the temperature sensor:2 to 75°C |
| K | Set the threshold of the smoke sensor:1 to 'High Density' |
| L | Set the threshold of the smoke sensor:1 to 'Medium Density' |
| M | Set the threshold of the smoke sensor:1 to 'Low Density' |
| N | Set the threshold of the smoke sensor:2 to 'High Density' |
| O | Set the threshold of the smoke sensor:2 to 'Medium Density' |
| P | Set the threshold of the smoke sensor:2 to 'Low Density' |
| R | Reset the System after an emergency situation. |

TABLE IV. LIST OF COMMAND CHARACTERS WITH CORRESPONDING ACTIONS

The software program checks the command character and the corresponding action is performed by the system consequently.

## VII. CONLCUSION AND FUTURE WORK

In this paper, I have given a clear view of an intelligent fire alarming system using wireless mobile communication. In my system the threshold of the sensor is configurable by the user. This feature has established the usability of the system in any country whatever the environmental temperature be. The user also has the provision to use it as an 'Early Fire Alert System' by setting the threshold of the temperature and smoke sensors to a lower level. So the system is capable of providing the user sufficient flexibility to entirely avoid any emergency situation due to fire blow. The incorporation of multiple sensors has provided the usability of the system not only in home but also in industries.

In case of any emergency situation due to fire blow, the SMS text is sent to a number of users. In many cases, a user may not notice this SMS and thus after sending SMS, CALL is also made from the server mobile. The user has the provision to set from the user interface of the program, developed for the server computer, that to how many people and to whom the SMS text will be sent and the CALL will be made. In the current system, no recorded voice is played when a phone call is made on the occurance of fire blow. But the user can easily get informed of the fire blow because from a specific mobile number that is known to user the call is made. In this case, a more sophisticated feature of playing a recorded voice when a call is made will be added to the system in future. And this feature can be used to make an intelligent call to fire brigade service in case of emergency situation.

Two stage microcontroller-connections will be developed in future to let more sensors to be connected to the system. In this case there will be one master microcontroller and other slave microcontrollers. Sensors will be connected to all the slave and master microcontrollers and when the slave microcontroller will receive an analog signal from a sensor surpassing the threshold for that sensor, it will send a signal to the master microcontroller through a non-ADC port and the master microcontroller will send the alert consequently. So the master microcontroller will check its ADC ports and the non-ADC ports to which the slave microcontrollers are connected for sending an alert. Hence the system, I will develop in future, will cover a larger area thus increasing the usability of the system to a great extent.

## REFERENCES

[1] Bibin John, "My Experience In Programming Microcontroller Using WinAvr/AvrGcc".

[2] Bibin John, "My Experience In Parallel Port Programming".

[3] Atmega32 datasheet.

[4] LM35 datasheet.

[5] Max232 datasheet.

[6] TTL Handbook

[7] www.atmel.com

[8] www.avrfreaks.net

[9] www.developershome.com

[10] www.embedtronics.com

[11] www.nokia.com

[12] www.smssolutions.net/tutorials/gsm/

[13] www.wikipedia.com